\documentclass[a4paper,11pt]{article}
\usepackage{pos}
\usepackage[compat=1.0.0]{tikz-feynman}

\tikzfeynmanset{
	fermion2/.style={
		/tikz/postaction={
			/tikz/decorate=false,
		},
	},
}
\tikzfeynmanset{
	fermion3/.style={
		/tikz/decoration={
			markings,
			mark=at position 0.5 with {
				\node[
				dot,
				fill,
				draw=none,
				] { };
			},
		},
		/tikz/postaction={
			/tikz/decorate=true,
		},
	},
}

\DeclareMathOperator*{\SumInt}{%
	\mathchoice%
	{\ooalign{\raisebox{.15\height}{\scalebox{0.9}{$\textstyle\sum$}}\cr\hidewidth$\displaystyle\int$\hidewidth\cr}}
	{\ooalign{\raisebox{.14\height}{\scalebox{.7}{$\textstyle\sum$}}\cr\hidewidth$\textstyle\int$\hidewidth\cr}}
	{\ooalign{\raisebox{.2\height}{\scalebox{.6}{$\scriptstyle\sum$}}\cr$\scriptstyle\int$\cr}}
	{\ooalign{\raisebox{.2\height}{\scalebox{.6}{$\scriptstyle\sum$}}\cr$\scriptstyle\int$\cr}}
}

\title{Crystalline Phases in QCD}

\author[a,b,c]{Theo F. Motta}

\affiliation[a]{Institut für Theoretische Physik,  Justus-Liebig-Universität Gie\ss en,
	35392 Gie\ss en, Germany}

\affiliation[b]{Technische Universität Darmstadt, Fachbereich Physik, Institut für Kernphysik,  Theoriezentrum, Schlossgartenstr.~2, 64289
	Darmstadt, Germany}

\affiliation[c]{Instituto de Física Teórica, Universidade Estadual Paulista, 01140-070 São Paulo, SP, Brazil}

\emailAdd{theo.motta@unesp.br}

\abstract{For decades now, low-energy models of QCD have shown indications that a crystalline quark phase could be stable at high chemical potentials. Beyond models, however, there are numerous difficulties in investigating such a hypothesis in full QCD, such as the sign problem. Functional methods do not suffer from the sign problem, and thus, can access the high--$\mu$ side of the QCD phase diagram. The main tool used to look for signs of inhomogeneous/crystalline phases in low-energy models is the so-called ``stability analysis''. In this talk, I show how the standard stability analysis was generalised to be applicable in any theory, including QCD.}

\FullConference{The XVIth Quark Confinement and the Hadron Spectrum Conference (QCHSC24)\\
 19-24 August, 2024\\
 Cairns Convention Centre, Cairns, Queensland, Australia\\}

\begin{document}
\maketitle

\section{Introduction}
When the order parameter for some symmetry becomes space-dependent, we say that this symmetry is broken \textit{inhomogeneously}. At large densities and low temperatures, chiral symmetry can be broken in such a way, as has been shown abundantly in low-energy models of strong interactions \cite{Buballa:2014tba,Nickel:2009wj,Thies:2006ti,Tripolt:2017zgc}. In full QCD, however, it is a rather tricky thing to investigate. The region where this is expected to happen is well beyond the current limit of applicability of lattice QCD and well within the energy bounds of non-perturbative phenomena. Within this region of the phase diagram, the only applicable tools are truncations of QCD via functional methods. In Ref.~\cite{Muller:2013tya}, the first solution of the Dyson-Schwinger Equations (DSEs) for an inhomogeneous phase was obtained and indeed such phases were found. Since inhomogeneous phases break translational symmetry, the quark propagator becomes a matrix in momentum space  $S(p_\text{in},p_\text{out})$ and it is very difficult to write (and solve) the quark DSE for a given ``shape'' of the quark condensate. Ideally, one would hope that we can take the exciting evidence shown in \cite{Muller:2013tya} and pursue the topic further, refining the DSE truncation, and see how robust is this evidence. This is in principle possible, however, it would require putting serious resources into this research and, if on the one hand, the model results and the results of \cite{Muller:2013tya} gave us good indications that this might indeed happen in QCD, on the other hand, Refs.~\cite{Pannullo:2021edr,Pannullo:2022eqh,Pannullo:2023one,Pannullo:2024sov,Winstel:2022jkk} showed that the model results could be somewhat artificial and Refs.~\cite{Pisarski:2018bct,Pisarski:2020dnx,Winstel:2024qle}
showed that including some types of fluctuations (disregarded in mean-field calculations and the truncation of Ref.~\cite{Muller:2013tya}) might wash out these non-trivial order parameters back into something spatially trivial. Although, if inhomogeneous phases are washed out by quantum fluctuations the remainder phase is not entirely boring. A necessary condition for inhomogeneous phases to appear is the presence of a moat-regime \cite{Rennecke:2021ovl}, where the mesonic dispersion relations are non-monotonic. This implies some sort of oscillatory pattern in the mesonic two-point function which is only dampened by the fluctuations, but not entirely washed out.

Patterned mesonic two-point functions, i.e. moat-regimes, would leave experimental signatures \cite{Pisarski:2021qof,Fukushima:2023tpv,Rennecke:2021ovl,Rennecke:2023xhc,Nussinov:2024erh}, whether they are dampened or infinitely long-range as in a truly inhomogeneos phase. 
Naturally, further work has to be put into QCD functional approaches to finally determine whether or not these phases exist. Rather than pursuing direct solutions of the DSE in an inhomogeneous phase, an alternative approach is to perform a stability analysis of the homogeneous solutions against small inhomogeneous perturbations. 

\section{Stability Analysis}

This is indeed one of the standard techniques used to look for these phases in QCD-inspired models. 
If one can calculate the free-energy of the theory as a function of the condensates $\phi$, by expanding the free-energy around the spatially trivial phase\footnote{Here we introduce the notation used in the rest of this proceedings, the bar $\bar\phi$ denotes homogeneous, and the delta $\delta\phi$ denotes the difference between the homogeneous and inhomogeneous $\delta\phi(x) = \phi(x) - \bar\phi$.}
\begin{equation}
    \phi(x) = \bar{\phi} + \delta\phi(x),
\end{equation}
one can compute the leading order contribution to the free energy $\Omega[\phi(x)]\approx\bar\Omega[\bar\phi] + \delta\Omega[\bar\phi,\delta\phi(x)]$ and, if $\delta\Omega$ turns out to be \textit{negative}, then the introduction of the inhomogeneous perturbation $\delta\phi(x)$ can \textit{lower} the free-energy and the homogeneous solution $\bar\phi$ is \textit{unstable}. In models with a local self-energy, such as the quark-meson model in mean field approximation, the NJL model, the Gross-Neveu model, etc, it turns out that $\delta\Omega$ is independent of $\delta\phi(x)$, i.e. $\delta\Omega[\bar\phi,\delta\phi(x)]=\delta\Omega[\bar\phi]$. One can then determine the stability of homogeneous phases easily, cleanly, and in a ``shape-independent'' way. In QCD, things are not so simple.

\subsection{QCD}

\newcommand{\Tr}{\text{Tr}}

Unfortunately, in QCD the free-energy cannot be written as a functional of the condensate and the quark self-energy is not local. Therefore, the stability analysis in QCD is not as easy, not as clean and cannot be done in a ``shape-agnostic'' way. It is nevertheless possible and can give us important insight. In Refs.~\cite{Motta:2023pks,Motta:2024agi} we managed to adapt this technique to be applied in any truncation of the homogeneous DSEs. We base the analysis on the so-called $n$PI formalism \cite{berges2004n,carrington2015techniques} where we can write the effective action (which is little more than an overall factor away from the free-energy) as a functional of the n-point functions of the theory. For instance, the 2PI effective action of QCD can be written as (ignoring 1-point functions)
\begin{equation}\label{eq:2PI}
	\begin{aligned}
		\Gamma[S,D,\Delta]&=\Tr\log\left[S^{-1}S_0\right] + \Tr\left[S_0^{-1}S\right] && \text{\color{darkgray}$\rightarrow$ Quark ``kinetic'' part}
		\\ &-\frac{1}{2}\Big(\Tr\log\left[D^{-1}D_0\right] 
		+ \Tr\left[D_0^{-1}D\right]\Big) && \text{\color{darkgray}$\rightarrow$ Gluon ``kinetic'' part}
		\\ &+\Tr\log\left[\Delta^{-1}\Delta_0\right] + \Tr\left[\Delta_0^{-1}\Delta\right] && \text{\color{darkgray}$\rightarrow$ Ghost ``kinetic'' part}
		\\ &+\Phi[S,D,\Delta] && \text{\color{darkgray}$\rightarrow$ Interaction part}
	\end{aligned}
\end{equation}
where $S$ is the quark propagator, $D$ the gluon propagator, and $\Delta$ the ghost propagator. The functional $\Phi$ contains the interactions and it can be written in a non-perturbative loop expansion. Since the point here is to construct the formalism of the stability analysis, we can take a strong simplification of the system. Consider quarks interacting via a frozen potential which models the yang-mills sector. That is, our effective action reads
\begin{equation}\label{CJT}
    \Gamma[S]=\Tr\log\left[S^{-1}S_0\right] + \Tr\left[S_0^{-1}S\right] +\frac{1}{2}
    \raisebox{-0.8cm}{
        \begin{tikzpicture}
            \begin{feynman}
                \vertex (a);
                \vertex [right=of a] (b);
                \vertex [right=of a] (b);
                \vertex [right=0.75cm of a] (l);
                \diagram*{
                    (a) -- [fermion3,half left] (b);
                    (b) -- [boson] (a);
                    (b) -- [fermion3,half left] (a);
                };\draw (l) node [gray, dot];
            \end{feynman}
        \end{tikzpicture}
    },
\end{equation}
where for an interaction functional we took a simple two-loop order truncation. Within this formalism we can easily write the homogeneous DSE for the quark propagator by finding the stationary points in the standard way, leading to
\begin{equation}\label{qDSE}
\centering%
\begin{tikzpicture}
    \begin{feynman}
        \vertex (a);
        \vertex [right=of a] (c);
        \vertex [right=0.5cm of c] (d);
        \vertex [above=0.4cm of c] (m1);
        \diagram*{
            (a) -- [fermion3] (c)
        };
        \draw (d) node { \({=}\)};
        \draw (m1) node { \(^{-1}\)};
        \vertex [right=0.5cm of d] (a1);
        \vertex [right=of a1] (c1);
        \vertex [right=0.5cm of c1] (d1);
        \vertex [above=0.4cm of c1] (m12);
        \diagram*{
            (a1) -- [fermion2] (c1)
        };
        \draw (d1) node { \({+}\)};
        \draw (m12) node { \(^{-1}\)};
        \vertex [right=0.5cm of d1] (a2);
        \vertex [right=0.55cm of a2] (gl1);
        \vertex [right=1cm of a2] (b2);
        \vertex [above=0.4cm of b2] (gldot);
        \vertex [right=1cm of b2] (c2);
        \vertex [left=0.5cm of c2] (gl2);
        \diagram*{
            (a2) -- [fermion2] (b2) -- [fermion2] (c2);
            (gl1) -- [boson,half left] (gl2)
        };
        \draw (gldot) node [gray, dot];
        \draw (b2) node [dot];
        \draw (gl2) node [];
    \end{feynman}
\end{tikzpicture}
\end{equation}
where the black-dotted propagators are fully dressed quark propagators. Once we obtain the homogeneous solution to Eq.~(\ref{qDSE}) we can then proceed to expand the effective action as follows
\begin{equation}
    S(k_1,k_2) = \bar{S}(k_1)\delta(k_1-k_2) + \delta S(k_1,k_2).
\end{equation}
where $\bar S$ is not only homogeneous, but also the \textit{chiral} homogeneous solution. This will become clearer soon, but suffice to say that choosing to expand around the chiral solution, we can use its stability boundary as an additional constraint to help with some technical issues to be discussed shortly.
Note that for the inhomogeneous part, the momentum is not conserved, due to translational symmetry breaking. We can then find the free-energy contribution by using the relation $\Omega = -(T/V) \Gamma$. The first non-trivial contribution to $\Omega$ from adding the perturbation $\delta S$ is quadratic in $\delta S$ and so it is labelled $\Omega^{(2)}$. One can easily show it to be
\begin{equation}\label{SC1}
  \begin{aligned}
    \Omega^{(2)}[\delta S]=
    &-\Tr\left[
    \bar{S}^{-1}(k_1)
    \delta S(k_1,k_2)
    \bar{S}^{-1}(k_2)
    \delta S(k_2,k_1)
    \right]\\
    &-g^2
Z_{1F}\text{Tr}
    \left[
    \gamma_\mu t^a  \delta S(k_1,k_2) \gamma_\nu t^b
    \delta S(k_2-q,k_1-q) D_{\mu\nu}^{ab}(q)\Gamma_{qg}(q)
    \right]\, ,
  \end{aligned}
\end{equation}
where $g$ is the coupling strength and $Z_{1F}$ is the vertex renormalization factor.
Here, one can already note how the non-locality of the quark-quark potential interferes with the calculation. With a local self-energy, both terms in $\Omega^{(2)}$ would contain one factor of $\delta S(k_1,k_2)$ and one of $\delta S(k_2,k_1)$ which could, under some assumptions on $\delta S$, be factored out as a positive factor of $|\delta S(k_1,k_2)|^2$ which would have no influence over the sign of $\Omega^{(2)}$. Since $D_{\mu\nu}^{ab}(q)\Gamma_{qg}(q)$ are momentum dependent, they shift one of the loop quark momenta and ruin this property.

Therefore, what we can do with this analysis is to ``test'' the stability of $\bar S$ against some specific shapes of $\delta S$, the perturbations. We call these the ``test-functions''. One is not, however, completely free to chose any test-function one would like to. In fact, not only there are several restrictions on the test-function, it is reasonable to chose something with physical meaning. Therefore, we chose to perform the following two variable shifts. By employing the Dyson series, one can write $\delta S$ as a test-function not on the propagator directly, but rather on the self-energy of the quark,
\begin{equation} \label{eq:stabilityII}
	\delta S(k_1,k_2)=\bar{S}(k_1)\delta \Sigma(k_1,k_2)\bar{S}(k_2),
\end{equation}
and the inhomogeneous perturbation to the self-energy can be written as follows
\begin{equation}\label{test-functionSig}
	\delta \Sigma(k_1,k_2) =\left(\frac{\delta m(k_1)+\delta m(k_2)}{2}\right)F(k_1-k_2),
\end{equation}
where $\delta m$ is a mass-like term.
The first benefit of this is that we can factor out all of the truly inhomogeneous part into one function $F$. Note that, if $F$ is taken to be $F(k_1-k_2) = \delta^{(4)}(k_1-k_2)$, then this becomes
\begin{equation}\label{test-functionSig2}
	\delta \Sigma(k_1,k_2) = \delta m(k_1) \delta^{(4)}(k_1-k_2)
\end{equation}
which is a homogeneous momentum-dependent mass term $\delta m$. In fact, taking the $k_1=k_2$ limit, one can use this to test \textit{homogeneous} chiral symmetry breaking in QCD (see \cite{Motta:2024agi}). When $k_1\neq k_2$, however, with an arbitrary $F$ function, this breaks translational symmetry as well as chiral symmetry. In other words, if inhomogeneous chiral symmetry breaking is the breaking of chiral symmetry together with translational symmetry, then in Eq.~(\ref{test-functionSig}), $\delta m$ breaks chiral and $F$ breaks translational symmetry.
Happily, since $F$ depends only on the difference of the momenta, it is insensitive to the momentum shift in Eq.~(\ref{SC1}) and we can indeed be agnostic with respect to its actual form. With these considerations $\Omega^{(2)}$ becomes\footnote{Here we introduce the notation $\SumInt_k = \sum_{\omega_k}\int \frac{d^3k}{(2\pi)^3}$ where $\omega_k$ are the relevant Matsubara frequencies.}
\begin{equation}\label{SC3}
 \begin{aligned}
    \Omega^{(2)}[\delta \Sigma]=
    &-\SumInt_{k_1 k_2} |F(k_1-k_2)|^2\times\text{tr}\Bigg[
    \bar{S}(k_1)
    \left(\frac{\delta m(k_1)+\delta m(k_2)}{2}\right)
    \bar{S}(k_2)
    \left(\frac{\delta m(k_2)+\delta m(k_1)}{2}\right)
    \Bigg]\\
    &
    \begin{aligned}-g^2
Z_{1F}\SumInt_{k_1 k_2 q} |F(k_1-k_2)|^2\times
        \text{tr}\Bigg[&\gamma_\mu t^a  \bar{S}(k_1)
        \left(\frac{\delta m(k_1)+\delta m(k_2)}{2}\right)
        \bar{S}(k_2)
        \gamma_\nu t^b
        \\&\bar{S}(k_2-q)
        \left(\frac{\delta m(k_2-q)+\delta m(k_1-q)}{2}\right)
        \bar{S}(k_1-q) D_{\mu\nu}^{ab}(q)\Gamma_{qg}(q)	\Bigg].
    \end{aligned}
 \end{aligned}
\end{equation}
Let $d$ be equal to (half of) the momentum difference, i.e. $2d=k_1-k_2$, we can the write the above as
\begin{equation}\label{eq:Omt1}
    \Omega^{(2)} = \SumInt_d |F(2d)|^2 \times  \tilde \Omega^{(2)}(d).
\end{equation}
and positivity or negativity of $\Omega^{(2)}$ is governed \textit{only} by $\tilde \Omega^{(2)}(d)$. If $\tilde \Omega^{(2)}(d)$ is negative for \textit{any value of} $d$, we can coneive of an $F$ function that is sufficially large for that $d$ such that $\Omega^{(2)}$ is negative.

\subsection{Complex Issues}
We can be agnostic with respect to $F$ but not with respect to the remaining part of the test-function, in particular to $\delta m$. Naïvely, one would think we can simply guess a function which decays rapidly enough with energy and momentum, like, say, a Gaussian, and then vary it's parameters until we find the lowest value of $\Omega^{(2)}$. In other words, evoke a standard variational method to find the most unstable configurations. If the ``most unstable'' still gives a positive value of $\Omega^{(2)}$, then, it is certainly stable (with respect to a Gaussian $\delta m$ on a test-function shaped like Eqs.~(\ref{eq:stabilityII},\ref{test-functionSig}), of course). Unfortunately, this is not the case. Since the integrand of the effective action is complex, and, in general, so is the effective mass and so could be the test-function, one must find a complex saddle point in the test-function space. $\Omega^{(2)}$ is bonded from below with respect to the real part of $\delta m$, as one would expect, but it is unbounded from below with respect to the imaginary part. Therefore, we must perform a saddle-point variational method, and this becomes extremely expensive numerically. In Refs.~\cite{Motta:2023pks,Motta:2024agi} this is explained in more detail. For the purposes of this talk, it will suffice to explain the following:

The biggest issue is not even the computational cost. The biggest issue is that with a conventional variational method to find a real minimum/maximum we can prove that the error is always positive/negative, i.e., any test-function would give you a larger/lower value of $\Omega^{(2)}$ than the true minimum/maximum. This is not the case for a saddle point. In other words, the Rayleigh–Ritz theorem does not apply. Therefore, if we guess a test-function with few parameters, we could find false instabilities. In Ref.~\cite{Motta:2024agi} we found a way to get around this issue for a particular region of the phase diagram. Exactly on top of the second order chiral transition and on top of the left spinodal line the \textit{homogeneous} stability analysis should give us exactly zero $\Omega^{(2)}_\text{homog} \equiv 0$ and we can use this as an extra constraint to fix the test-function's parameters. Unfortunately, though, we had to restrict the analysis to this rather small region of the phase diagram. These difficulties notwithstanding, we performed the analysis for three different models of the running coupling.

\subsection{Running Coupling Models}
The remaining ingredient is how do we actually model the quark-quark potential. Quite simply, we take the gluon propagator
\begin{align}\label{eq:qProp}
D_{\mu\nu}^{ab}(q) &= \Bigg(P_{\mu\nu}^{T}(q) \,\frac{Z_{T}(q)}{q^2} + P_{\mu\nu}^{L}(q) \,\frac{Z_{L}(q)}{q^2} \Bigg)\delta^{ab} \,,
\end{align}
where $P^{T,L}$ are the projectors transverse and longitudinal to the heat-bath,
together with the vertex dressing function\footnote{Not to be confused with the effective action $\Gamma$.} $\Gamma(q)$, and define effective running couplings
\begin{equation}
    \alpha_{T,L}(q) = \frac{g^2}{4 \pi} \frac{Z_{1F}}{Z_2^2} \Gamma_{qg}(q) Z_{T,L}(q),
\end{equation}
which can be taken such that $\alpha/q^2$ has a Gaussian form (Qin-Chang model \cite{Qin:2011dd})
\begin{equation}
    \frac{\alpha_\text{QCIR}(q^2)}{q^2} = 
			\frac{2\pi}{\omega_\text{QC}^4} D e^{-q^2/\omega_\text{QC}^2},
\end{equation}
where $\omega_\text{QC}=600$~MeV and $D=1$~GeV$^2$ are model parameters, or a Maxwellian form (Maris-Tandy model \cite{Maris:1999nt})
\begin{equation}
    \frac{\alpha_\text{MT}(q^2)}{q^2} = 
			\pi \frac{\eta^7}{\Lambda_\text{MT}^4} q^2 e^{-\eta^2\,q^2/\Lambda_\text{MT}^2},
\end{equation}
where here $\eta = 1.8$ and $\Lambda_\text{MT}=720$ MeV are the model parameters.
These are models for the IR behaviour of the coupling. As for the UV part, usually one adds the 1-loop perturbative running coupling. Quite often one neglects this part, since we are mostly interested in infrared physics. However, for completeness, we calculate the Maris-Tandy model with and without the so-called UV ``log-tail''
\begin{equation}
    \alpha_\text{log-tail}(q^2) = 
			\frac{4\pi^2\gamma_m}{(1/2)\log(\tau+(1+q^2/\Lambda_\text{QCD}^2)^2)}
			\Big({1-e^{-q^2/4m_t^2}}\Big),
\end{equation}
where $\gamma_m=12/(33-2N_F)$, $N_F=4$, $m_t=500$ MeV, $\Lambda_\text{QCD}=234$ MeV and $\tau = e^2-1$.

\section{Results}

\begin{figure}
    \centering
    \includegraphics[width=0.48\linewidth]{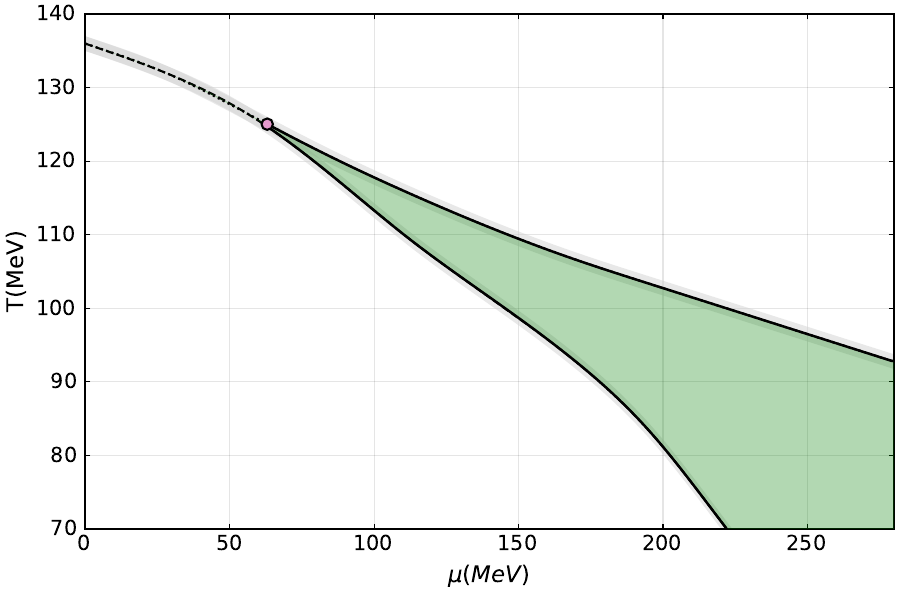}
    \includegraphics[width=0.48\linewidth]{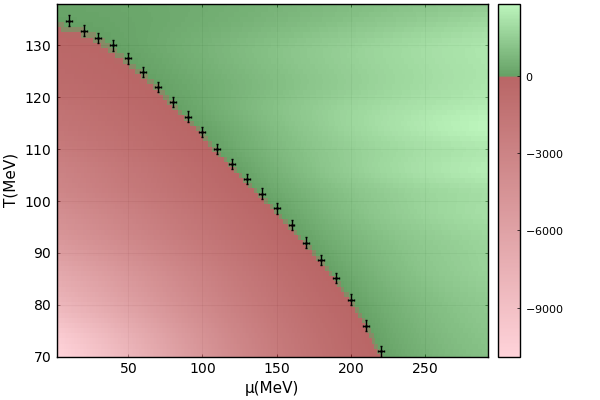}

    \includegraphics[width=0.48\linewidth]{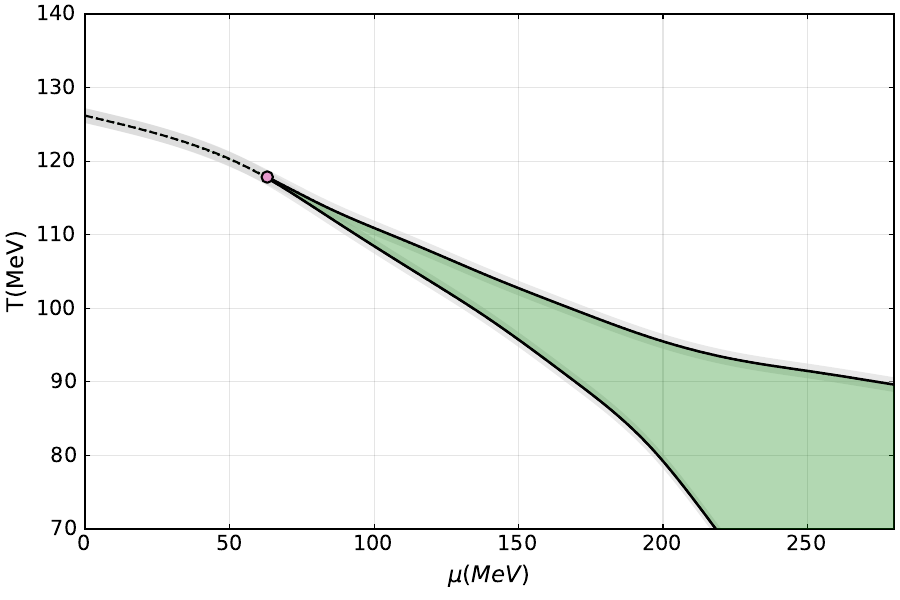}
    \includegraphics[width=0.48\linewidth]{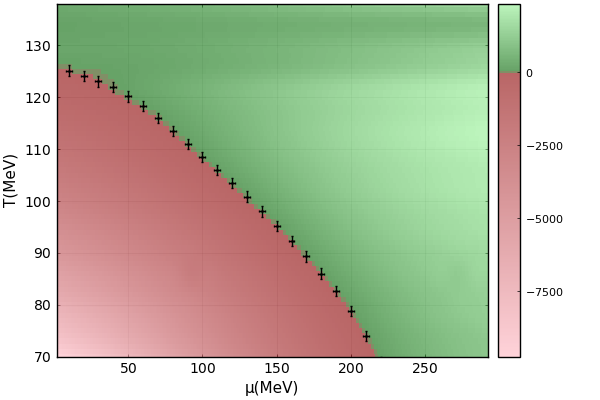}

    \includegraphics[width=0.48\linewidth]{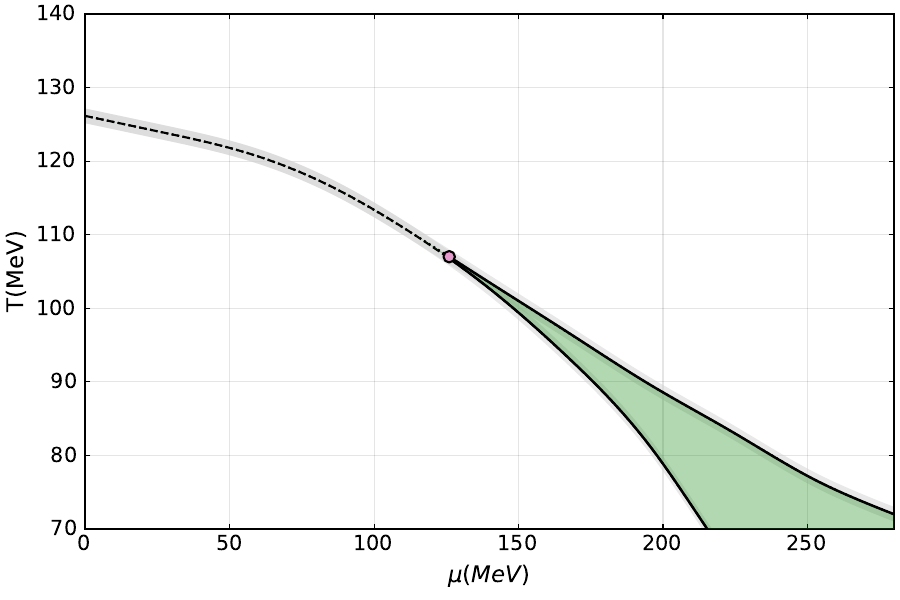}
    \includegraphics[width=0.48\linewidth]{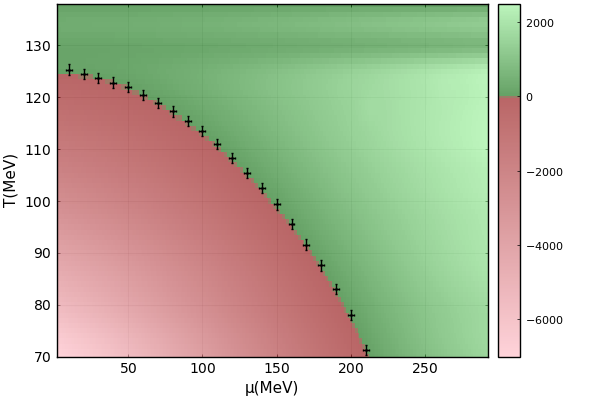}
    \caption{The first column of this plot shows the full homogeneous phase diagrams in the three models referred to in the text. the top-left plot corresponds to the full Maris-Tandy model (with log-tail), the centre-left is the Maris-Tandy model with only the infrared part, and the bottom-left uses the infrared part of the Qin-Chang model. On the right-hand-side the plots correspond to the same model as the plots on their left. The color-bar shows the result of $\tilde\Omega^{(2)}(d=0)$ and the black crosses show the second-order homogeneous chiral transition and the left spinodal lines, also shown in the plots on their left.}
    \label{fig:pds}
\end{figure}

In Ref.~\cite{Motta:2024agi} the following form for $\delta m$ was taken
\begin{equation}
    \delta m(\omega,\vec k) = e^{-\left(\frac{\omega^2}{L_1^2}+\frac{\vec k^2}{L_2^2}\right)}
		+ i L_3 \frac{\omega}{\omega_0} e^{-\left(\frac{\omega^2}{L_4^2}+\frac{\vec k^2}{L_5^2}\right)}
\end{equation}
and the parameters were fixed according to the complex saddle point condition. That is, we minimise $\Omega^{(2)}$ with respect to $L_1$ and $L_2$, the parameters for the real part, and we maximise with respect to $L_{3,4,5}$ which parametrise the imaginary part. We know this to be an exhaustive parameter set because this form reproduces the correct stability boundaries for the chiral solution. In Fig.~\ref{fig:pds} we show the phase diagrams for the three models mentioned above and, on the right of each diagram we show the result for the stability analysis assuming $d=0$, that is, the homogeneous stability analysis of the chiral solution, with respect to chiral symmetry breaking. Since the exact boundaries were reproduced, we decide the test function is appropriate to be used along the second order transition and the left spinodal. Once we perform the analysis for $d>0$, if the stability condition becomes negative in this region, for any finite $d$, this would mean that the chiral solution is also unstable with respect to translational symmetry breaking. 

\begin{figure}
    \centering
    \includegraphics[width=0.98\linewidth]{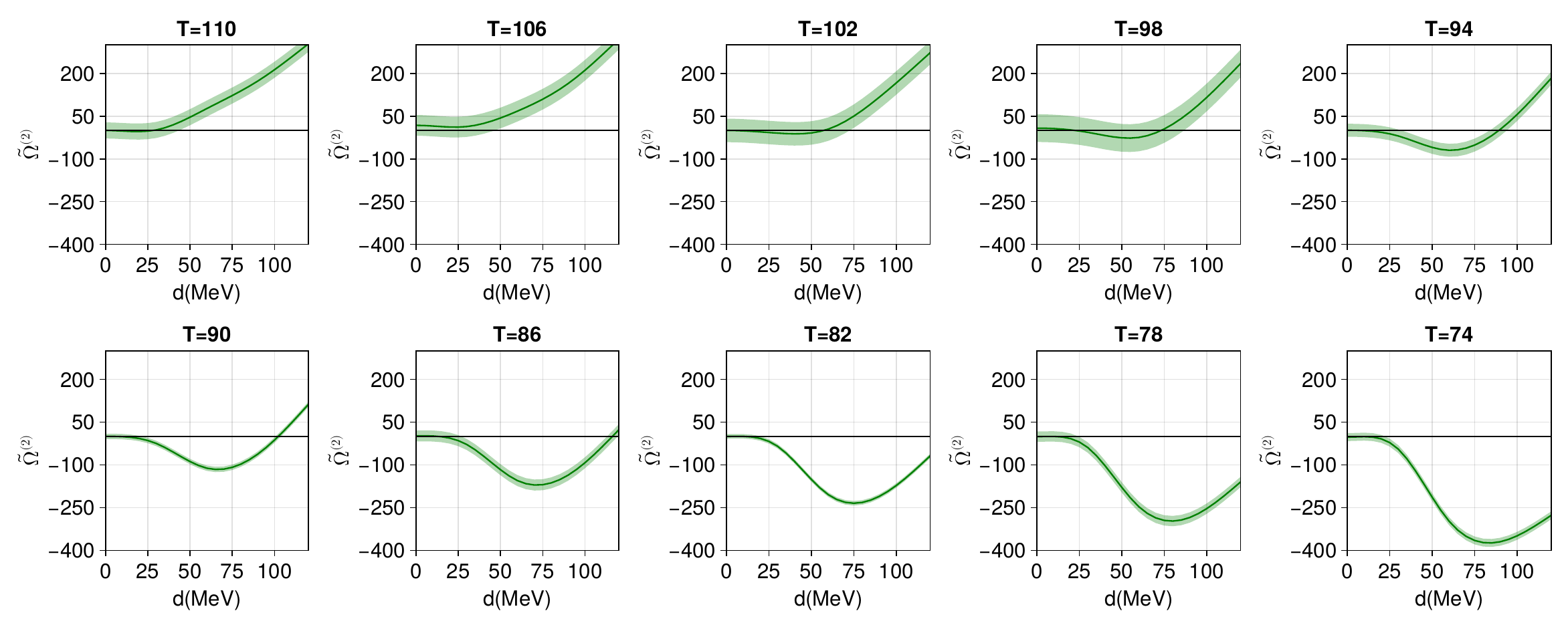}
    
    \includegraphics[width=0.48\linewidth]{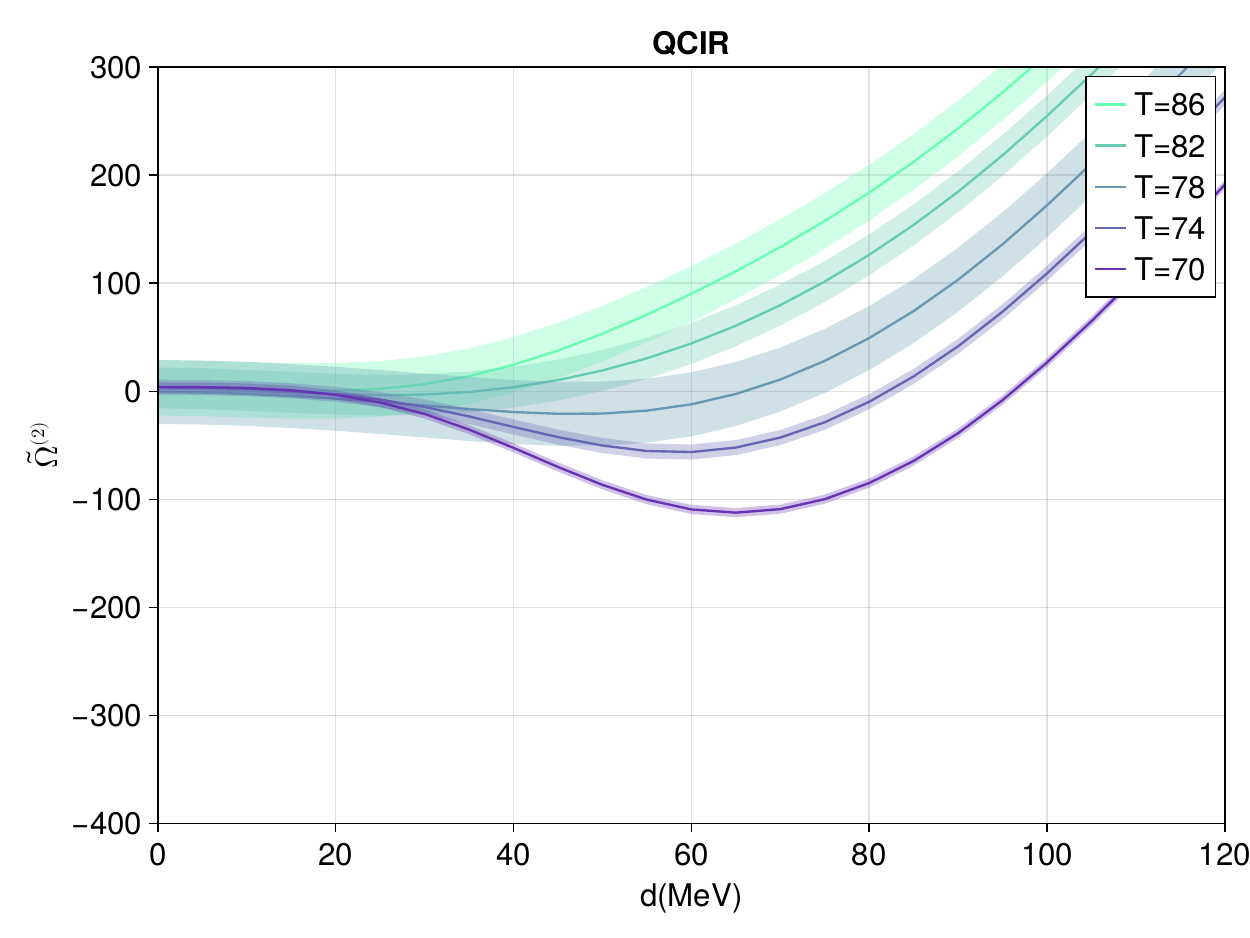}
    \includegraphics[width=0.48\linewidth]{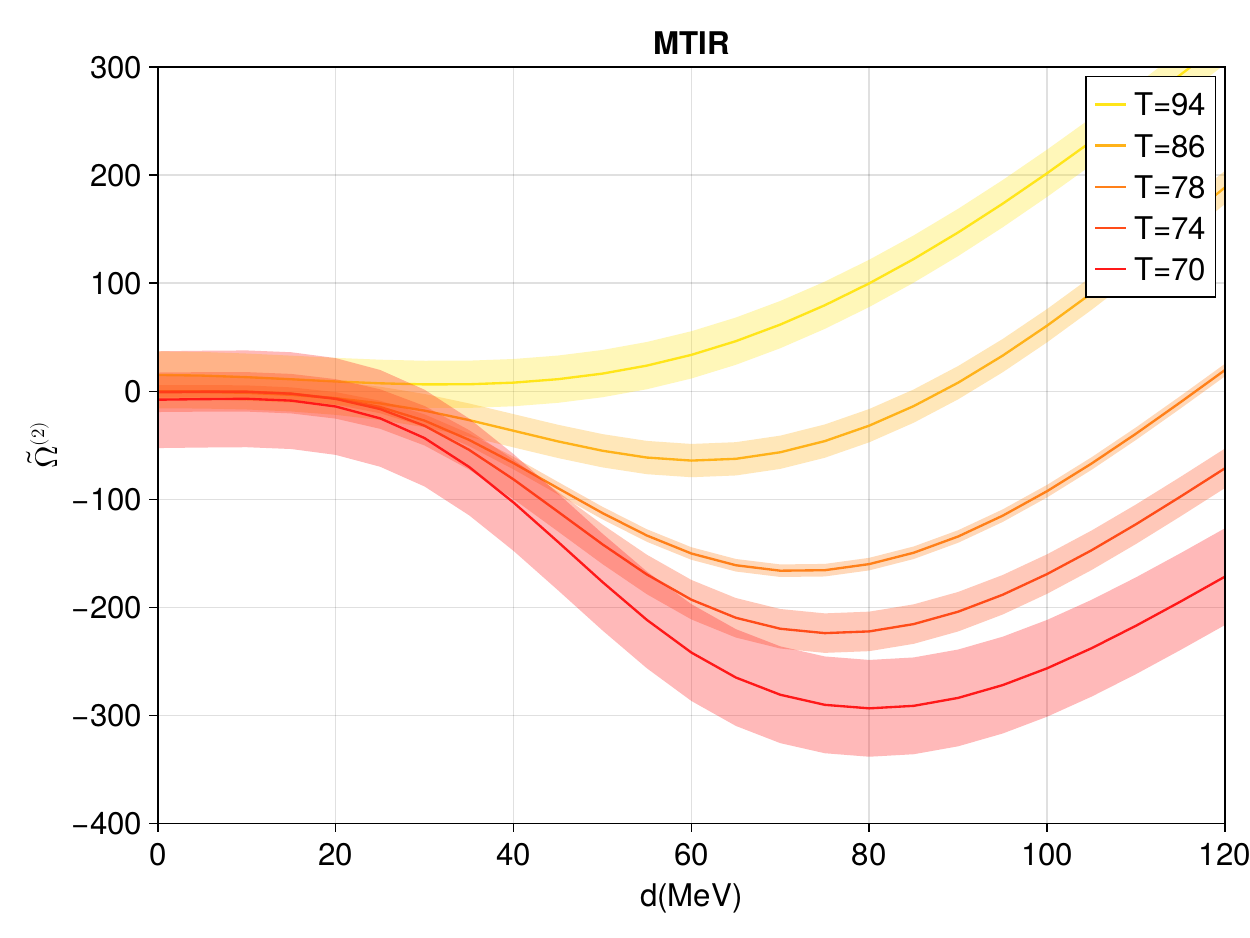}
    \caption{Here we show, for the same models as before, the value of $\tilde \Omega^{(2)}$ as a function of $q$ calculated in the line where $\tilde \Omega^{(2)}(0)=0$. The smaller plots on the top show results for the full Maris-Tandy model and the two plots below, labelled QCIR and MTIR show results for the Qin-Chang and Maris-Tandy models without the UV contribution.}
    \label{fig:slices}
\end{figure}

In Fig.~\ref{fig:slices} we show exactly that. The smaller plots on the top show results for the full Maris-Tandy model, i.e. Maris-Tandy with the perturbative log-tail. Each plot is calculated in a different point of the homogeneous stability boundary of the chiral solution. In other words, each plot is calculated along the line of zero $\Omega^{(2)}$ in the right-hand-side plots in Fig.~\ref{fig:pds}. There, we can clearly see that, as the temperatures get lower, we do get negative values of $\Omega^{(2)}$ for finite $d$. The same is found in the IR-only models, shown below. This is a clear and consistent pattern. The chiral solution is unstable with respect to inhomogeneous chiral symmetry breaking in this region. 

\section{Conclusions and Subsequent Work}

\begin{figure}
    \centering
    \includegraphics[width=0.48\linewidth]{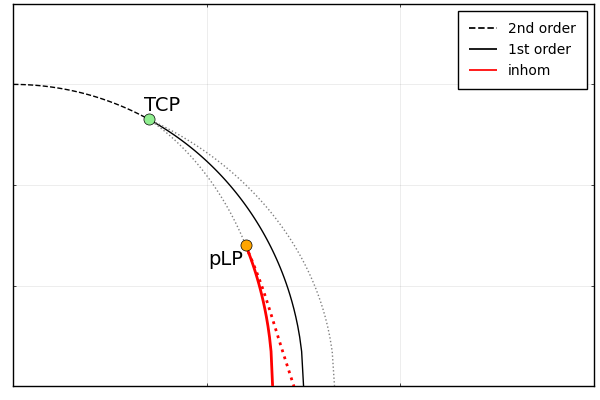}
    \includegraphics[width=0.48\linewidth]{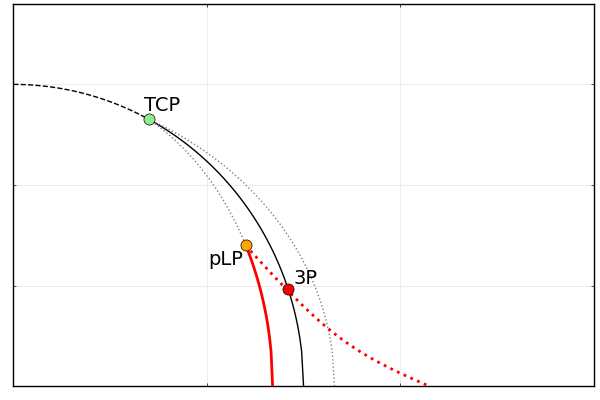}
    \caption{Sketches of the full phase diagram shown in Ref.~\cite{Motta:2024agi} where TCP stands for Tri-Critical-Point, pLP stands for proto-Lifschitz Point, and 3P, naturally, is a triple-point.}
    \label{fig:sketches}
\end{figure}

It is clear that there is a tendency for the chiral solution to break both chiral symmetry and translational symmetry in the region we analysed. However, the obvious question is, does this have any physical meaning? The question is appropriate since the chiral solution in this region (below the TCP) is energetically disfavoured. Therefore, this shows an instability of a phase that is not the true ground-state and therefore this might not have any real effect. It is the case, however, that in QCD-inspired models, when an inhomogeneous phase is found, the chiral solution \textit{is} unstable in the same manner in the region where it is disfavored. However, this is far from a proof. We therefore speculate that the final phase diagram should look like one of the two plots in Fig.~\ref{fig:sketches}. Since the minimum of $\tilde \Omega^{(2)}$ in Fig.~\ref{fig:slices} gets deeper and deeper for lower temperatures, we expect that the complete instability region (away from the left spinodal) would extend further in $\mu$ for lower temperatures. However, if this extension does \textit{not} cross the first order transition line, the instability is meaningless, as depicted in the left hand plot of Fig.~\ref{fig:sketches}. If it \textit{does} cross the first-order homogeneous transition, then this instability would persist on a region where the chiral solution not only is favoured, it is the \textit{only} homogeneous solution. Up until the publication of Ref.~\cite{Motta:2024agi} and the delivery of this talk, it was impossible for us to say which of the two scenarios is truly realised. Now, with the work done and released in Ref.~\cite{Motta:2024rvk}, which is based in a completely new method that bypasses the need for a saddle-point search, we were able to confirm that the diagram on the right of Fig.~\ref{fig:sketches} is the one that correspond to the reality of these models. In Ref.~\cite{Motta:2024rvk} we also confirm the location of the proto-Lifshcitz Points (pLP) in Fig.~\ref{fig:sketches} which is basically a Lifschitz point for some phase that stays inside a region where the phase is disfavored. This point corresponds to the region where, in Fig.~\ref{fig:slices}, $\Omega^{(2)}$ starts to become negative.

With the work done in Ref.~\cite{Motta:2024rvk} which fully confirms the findings shown here, the two methods can be seen as complementary. This work is completely applicable to any nPI truncation of QCD and future work will show whether or not we have an instability towards inhomogeneous chiral symmetry breaking in a closer-to-QCD truncation.

\section*{Acknowledgments}
This work has been supported by the Alexander von Humboldt Foundation, and by the Funda\c{c}\~ao de Amparo \`a Pesquisa do Estado de S\~ao Paulo (FAPESP), grant 2024/13426-0.

\bibliographystyle{ieeetr}
\bibliography{stabilitybib.bib}

\end{document}